\documentclass[aps,preprint,showpacs,groupedaddress]{revtex4-1}
\usepackage{amssymb}
\usepackage{mathrsfs}
\usepackage{pstricks}
\usepackage{color}
\usepackage{graphicx}
\usepackage[fleqn]{amsmath}
\usepackage{slashed}
\usepackage{amsmath}
\usepackage{mathtools}
\usepackage{array}
\usepackage{tabularx}
\usepackage{physics}
\usepackage{slashed}
\usepackage[utf8]{inputenc}
\usepackage[T1]{fontenc}
\usepackage{float}

\begin{document}


\title{\bf Is Kaniadakis $\kappa$-generalized statistical mechanics general?}


\author{T. F. A. Alves}
\author{J. F. da Silva Neto}
\author{F. W. S. Lima}
\author{Paulo R. S. Carvalho}
\email{prscarvalho@ufpi.edu.br}
\affiliation{\it Departamento de F\'\i sica, Universidade Federal do Piau\'\i, 64049-550, Teresina, PI, Brazil}
\author{G. A. Alves}
\affiliation{\it Departamento de F\'{i}sica, Universidade Estadual do Piau\'{i}, 64002-150, Teresina - PI, Brazil}





\begin{abstract}
In this Letter we introduce some field-theoretic approach for computing the critical properties of systems undergoing continuous phase transitions governed by the $\kappa$-generalized statistics, namely $\kappa$-generalized statistical field theory. In particular, we show, by computations through analytic and simulation results, that the $\kappa$-generalized Ising-like systems are not capable of describing the nonconventional critical properties of real imperfect crystals, \emph{e. g.} of manganites, as some alternative generalized theory is, namely nonextensive statistical field theory, as shown recently in literature. Although $\kappa$-Ising-like systems do not depend on $\kappa$, we show that a few distinct systems do. Thus the $\kappa$-generalized statistical field theory is not general, \emph{i. e.} it fails to generalize Ising-like systems for describing the critical behavior of imperfect crystals, and must be discarded as one generalizing statistical mechanics. For the latter systems we present the physical interpretation of the theory by furnishing the general physical interpretation of the deformation $\kappa$-parameter.
\end{abstract}


\maketitle


\section{Introduction}

\par Motivated by the incapacity of describing some physical phenomena \cite{PhysRevLett.115.238301,Lutz2013,PhysRevD.91.114027,TSALLIS1998534,TSALLIS1999,TSALLIS2021} through Boltzmann-Gibbs (BG) statistics, some attempts for generalizing that statistics were made. In fact, some generalized statistics were proposed \cite{Tsallis1988,KANIADAKIS2001405,BECK2003267,PhysRevE.71.046128,Eur.Phys.J.B70.3}. In the generalization process, only that statistics satisfying a set of consistency conditions will survive. Among these conditions, the consistent statistics have to be obtained from a maximum principle and a trace-form entropy. Other consistency requirements are positivity, continuity, symmetry, expansibility, decisivity, maximality, concavity and Lesche stability. Another reasonable condition is its applicability to all problems to be generalized. Suppose that there is only and only one experimental situation in which for being described we need a generalized statistics and this statistics is not capable of describing such a situation. Then that statistics is not general and must be discarded as one trying to generalize statistical mechanics. In this direction, we can desire to generalize one of the most fundamental applications of BG statistical mechanics, that of the computation of the critical properties of continuous phase transitions \cite{Stanley}. For obtaining some of these properties, \emph{e. g.} critical exponents, Kenneth Wilson developed the field-theoretic renormalization group \cite{PhysRevLett.28.240,PhysRevLett.28.548}. Such a mathematical tool was successfully applied and furnished precise values for the critical indices that showed a satisfactory agreement with experimental results \cite{Wilson197475}. So, in the intention of obtaining a generalized theory of phase transitions valid in a generalized realm, recently, some generalized field-theoretic approach was designed \cite{Submitted}, namely nonextensive statistical field theory (NSFT). In this generalized approach, some new $q$-parameter is introduced in the generalization process and the resulting theory is valid for $|1 - q| < 1$. It was physically interpreted as one encoding some effective interaction, which can be turned off in the limit $q\rightarrow 1$ thus recovering the BG results \cite{PhysRevLett.28.240,PhysRevLett.28.548}. Then a new generalized universality class arose, the O($N$)$_{q}$ one, from which emerged nonextensive Ising-like models as $q$-Ising ($N = 1$), $q$-XY ($N = 2$), $q$-Heisenberg ($N = 3$), -self-avoiding random walk ($N = 0$) and $q$-spherical models ($N \rightarrow \infty$). Some other $q$-generalized models are the $q$- percolation and Yang-Lee edge singularity, -$\phi^{6}$ theory, -long-range, -Gross-Neveu, -uniaxial systems with strong dipolar forces, -Lifshitz, -long-range $\phi^{3}$ theory, -$\phi^{2k}$  multicritical points of order $k$, -Gross-Neveu-Yukawa, -short- and -long-range directed and dynamic isotropic percolations \cite{Submitted}. Now, the nonconventional critical indices depend on the dimension $d$, $N$ and symmetry of some $N$-component order parameter and if the interactions present are of short- or long-range type and on $q$. It was shown that nonextensivity was associated only to small length scales fluctuations and its effects emerged from radiative loop corrections. It does not manifest at large length scales, once such large scales would be probed by a supposed nonextensive thermodynamics. But a nonextensive thermodynamics does not exist, as it is shown in Ref \cite{PhysRevLett.88.020601}. However, there is a full set of thermodynamical properties valid for $q\neq 1$: see, for instance \cite{PhysRevE.89.022117,Andrade_2014,PhysRevE.91.022135}. In fact, through a transformation of variables \cite{PhysRevLett.88.020601}, the supposed nonextensive thermodynamics can be mapped into its extensive counterpart. Also, the predictions of NSFT, through its Escort distribution version \cite{CARVALHO2022137147}, presented an excellent agreement with those obtained from computer simulations, within the margin of error, for the static and dynamic critical indices for the nonexetnsive version of the two-dimensional Ising model \cite{PhysRevE.102.012116}. The aim of this Letter is: $1$) Introducing the $\kappa$-generalized version of the Kenneth Wilson's field-theoretic renormalization group in momentum space \cite{PhysRevLett.28.240}, namely $\kappa$-Statistical Field Theory ($\kappa$-SFT) and $2$) Investigating what are the BG systems presenting the corresponding $\kappa$-generalized generalization, \emph{i. e.}, $\kappa$-generalized systems whose critical exponents depend on $\kappa$.

\section{$\kappa$-SFT}

\par We introduce the $\kappa$-SFT by defining its Euclidean generating functional as 
\begin{eqnarray}\label{huyhtrjisd}
Z[J] = \mathcal{N}^{-1}\exp_{\kappa}\left[-\int d^{d}x\mathcal{L}_{int}\left(\frac{\delta}{\delta J(x)}\right)\right]\int\exp\left[\frac{1}{2}\int d^{d}xd^{d}x^{\prime}J(x)G_{0}(x-x^{\prime})J(x^{\prime})\right],
\end{eqnarray}
where 
\begin{eqnarray}\label{jhglhjlkisd}
\exp_{\kappa}(-x) = \left(\sqrt{1 + \kappa^{2}x^{2}} - \kappa x\right)^{1/\kappa}
\end{eqnarray}
is the $\kappa$-generalized exponential function \cite{KANIADAKIS2001405} and $\kappa\in$ ($-1, 1$) and $G_{0}(x-x^{\prime})$ is the free propagator of the theory. Analogously to the Ref. [21] (where the corresponding second term is extensive,  once it is associated to the free propagator, which can be deﬁned only in the extensive scenario, we make $q = 1$ and obtain the conventional exponential), we make $\kappa = 0$ in the second term of Eq. (\ref{huyhtrjisd}) and obtain the conventional exponential once it is associated to the free propagator, which can be deﬁned only in the nongeneralized realm. The constant $\mathcal{N}$ is determined from $Z[J=0] = 1$. Now we study some $\kappa$-generalized models.

\section{$\kappa$-Ising-like systems}

\par By applying perturbation theory for some O($N$)-symmetric $N$-component self-interacting $\lambda\phi^{4}$ scalar field theory, we obtain the $\kappa$-generalized static, through six distinct and independent methods in dimensions $d = 4 - \epsilon$, and dynamic critical exponents as 
\begin{eqnarray}\label{etaphi4}
\eta_{\kappa} = \eta , \hspace{1cm} \nu_{\kappa} = \nu , \hspace{1cm} z_{\kappa} = z ,
\end{eqnarray}
where $\eta_{\kappa}$, $\nu_{\kappa}$ and $z_{\kappa}$ are the corresponding nongeneralized critical exponents valid for all loop levels. We observe that the aforementioned critical indices are the same as their nongeneralized counterparts \cite{Wilson197475}. This shows that for the corresponding Ising-like systems as $\kappa$-generalized Ising, XY, Heinsenberg, self-avoiding random walk  and spherical models, the associated critical exponents do not depend on $\kappa$. The same occurs for $\kappa$-generalized $\phi^{6}$ \cite{STEPHEN197389,PhysRevE.60.2071,Hager_2002}, long-range \cite{BrezinEandParisiGandRicci-TersenghiF,PhysRevLett.29.917}, Gross-Neveu \cite{PhysRevD.10.3235,PhysRevD.94.125028}, uniaxial strong dipolar forces \cite{PhysRevB.13.251}, spherical \cite{PhysRev.86.821,PhysRevLett.28.240}, Lifshitz \cite{PhysRevLett.35.1678,PhysRevB.67.104415,PhysRevB.72.224432,Albuquerque_2001,LEITE2004281,PhysRevB.61.14691,PhysRevB.68.052408,FARIAS,Borba,Santos_2014,deSena_2015,Santos_2019,Santos_20192,Leite_2022} and multicritical points of order $k$ \cite{C.ItzyksonJ.M.Drouffe}. Then, the $\kappa$-SFT is not suitable for describing the critical properties of nonconventional real imperfect crystals, for example of manganites \cite{Magnetochemistry.Turki,KHELIFI2014149,PhysRevB.75.024419,OMRI20123122,Ghosh_2005,doi:10.1063/1.2795796,GHODHBANE2013558,J.Appl.Phys.A.Berger,PhysRevB.68.144408,BenJemaa,PhysRevB.70.104417,PhysRevB.79.214426,J.Appl.Phys.Vasiliu-Doloca,YU2018393,PhysRevB.92.024409,ZHANG2013146,PHAN201440,RSCAdvJeddi,HCINI20152042,Phys.SolidStateBaazaoui} presenting defects, impurities, inhomogeneities, size of the clusters, random magnetic dilution, magnetocrystalline anisotropies etc. and the competition among them, as the nonextensive statistical field theory is as shown recently in literature \cite{Submitted}.

\par Now, we compare the field-theoretic results of this section with the ones obtained by the Monte-Carlo simulation of the 2D Ising Model. We consider a square lattice with $L^2$ nodes and periodic boundary conditions, where we can assign a system state
\begin{equation}
  \boldsymbol{\sigma} = \left( \sigma_1, \sigma_2, ..., \sigma_N \right),
\end{equation}
where $N=L^2$ and each stochastic spin variable can have the values $\sigma_{i} = \pm 1$. We can start the dynamics with a random configuration, and at each step, we randomly choose one spin to be updated. Then, we try a spin flip with the $\kappa$-generalized Metropolis rate
\begin{equation}
  w_i(\boldsymbol{\sigma}) = \text{Minimum}\left[ 1,\frac{\exp_k{\left(-e^{(2)}_i/T\right)}}{\exp_k{\left(-e^{(1)}_i/T\right)}}\right].
  \label{ising-metropolis-kaniadakis}
\end{equation}
where $T$ is the temperature and $e^{(1)}_i$ and $e^{(2)}_i$ are the energies of the spin $i$ before and after the spin-flip, respectively. The local spin energies $e^{(1)}_i$ and $e^{(2)}_i$ are given by the Ising Hamiltonian
\begin{equation}
  H = \sum_{\left<i,j\right>}^{L^2}\sigma_i\sigma_j
  \label{ising-hamiltonian}
\end{equation}
where the index $j$ runs on the first neighbors of node $i$. The following Master equation describes the Ising dynamics\cite{Landau-2015}
\begin{equation}
  \frac{d}{dt} \mathcal{P}_{\boldsymbol{\sigma}} = \sum_i^N w_i(\boldsymbol{\sigma^i}) \mathcal{P}_{\boldsymbol{\sigma^i}} - w_i(\boldsymbol{\sigma}) \mathcal{P}_{\boldsymbol{\sigma}},
  \label{masterequation}
\end{equation}
where $\mathcal{P}_{\boldsymbol{\sigma}}$ is the occupation probability of one system state and
\begin{equation}
  \boldsymbol{\sigma^i} = \left( \sigma_1, \sigma_2, ...,-\sigma_i, ..., \sigma_{N} \right),
\end{equation}
is the state after a spin-flip. The occupation probabilities should have a stationary solution where the local spin energies obey the Kaniadakis distribution if one chooses the rates $w_i$ in Eq.\ (\ref{ising-metropolis-kaniadakis}).

We define a Monte Carlo step as the sequential update of $L^2$ spins. We wait for the system to reach the stationary state from an initial random state by updating the system $N_\text{term}$ thermalization steps. When the system is in the stationary state, we begin to collect a time series with $N_t$ elements of the thermodynamic parameters, for example, the mean magnetization
\begin{equation}
  m_\ell = \left| \frac{1}{L^2} \sum_i \sigma_i(t_\ell) \right|,
\end{equation}
and the mean internal energy 
\begin{equation}
  e_\ell = \frac{1}{L^2} \left<H\right>(t_\ell),
\end{equation}
where $t_\ell$ ($\ell = 1, 2, ..., N_t$) is the simulation time after the thermalization, which is a multiple of the Monte Carlo step time because we also discard Monte-Carlo steps between two elements of the time series to avoid data correlation and critical slowing down effects \cite{Landau-2015}. The observable moments are then given by
\begin{eqnarray}
  \left<m^n\right> &=& \frac{1}{N_t} \sum_\ell^{N_t} m^n_\ell \nonumber \\
  \left<e^n\right> &=& \frac{1}{N_t} \sum_\ell^{N_t} e^n_\ell
\end{eqnarray}
And the following averages on the ensembles of $m_\ell$ and $e_\ell$ time series yield
\begin{eqnarray}
  U(T,L)    &=& 1 - \frac{\left<m^4\right>}{3\left<m^2\right>}, \nonumber \\
  M(T,L)    &=& \left< m \right>, \nonumber \\
  \chi(T,L) &=& \frac{L^2}{T} \left( \left<m^2\right> - \left<m\right>^2 \right), \nonumber \\
  c(T,L)    &=& \frac{L^2}{T^2}\left( \left<e^2\right> - \left<e\right>^2 \right),
  \label{ising-observables}
\end{eqnarray}
which are the Binder cumulant, the magnetization, the magnetic susceptibility, and the specific heat, respectively. The Binder cumulant should not depend on the system size in the critical temperature allowing for an estimate of the critical temperature where the curves for different lattice sizes approximately cross \cite{Landau-2015}. In addition, from resampling the time series, one can estimate error bars \cite{Landau-2015}. The above thermodynamic quantities in Eq.\ (\ref{ising-observables}) should scale as the finite system size in two dimensions as
\begin{eqnarray}
  U(T,L)    &\propto&                              F_U    \left[ L^{-1/\nu_\kappa}\left(T - T_c\right) \right] \nonumber \\
  M(T,L)    &\propto& L^{-\beta_\kappa/\nu_\kappa} F_M    \left[ L^{-1/\nu_\kappa}\left(T - T_c\right) \right] \nonumber \\
  \chi(T,L) &\propto& L^{\gamma_\kappa/\nu_\kappa} F_\chi \left[ L^{-1/\nu_\kappa}\left(T - T_c\right) \right] \nonumber \\
  c(T,L)    &\propto& L^{\alpha_\kappa/\nu_\kappa} F_c    \left[ L^{-1/\nu_\kappa}\left(T - T_c\right) \right]
  \label{scaling-functions-ising-2d}
\end{eqnarray}
close to the critical temperature $T_c$ in a continuous phase transition, respectively, where $F_U$, $F_M$, $F_\chi$, and $F_c$ are scaling functions.

\begin{figure*}[!ht]
\begin{center}
\includegraphics[scale=0.125]{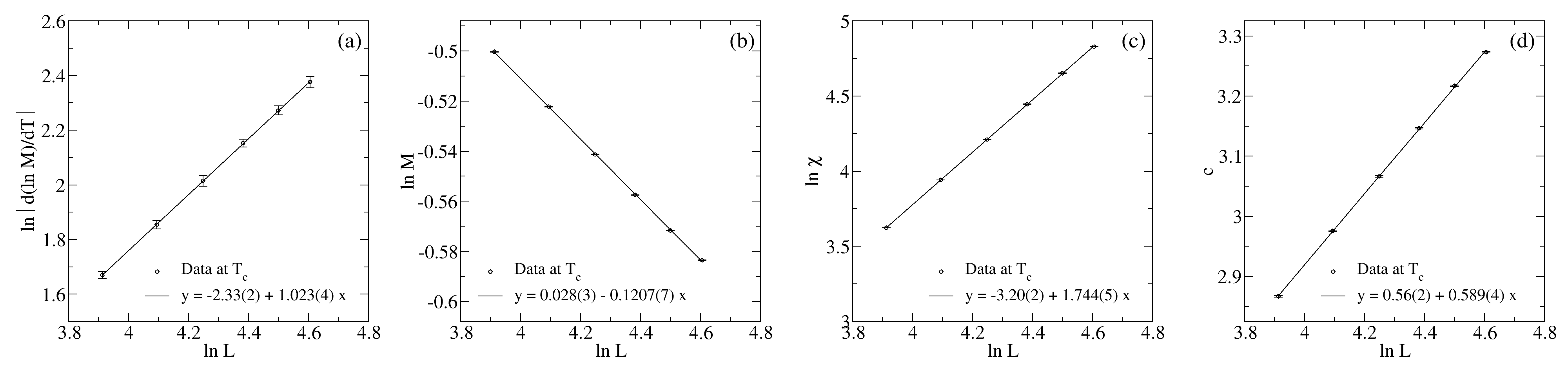}
\end{center}
\caption{We show simulation results of $d\left(\ln M\right)/dT$, $M$, $\chi$ and $c$ as functions of the system size in panels (a), (b), (c), and (d), respectively, obtained from the Monte-Carlo simulation of the 2D Ising model with the $\kappa$-generalized Metropolis rate in Eq.\ (\ref{ising-metropolis-kaniadakis}) close to the critical temperature $T_c \approx 1.728(5)$. We note that panels (a), (b), and (c) are double log plots, and panel (d) is a semilog plot. The data slopes in the double log plots at panels (a), (b), and (c) yield estimates of the respective critical exponent ratios $1/\nu_\kappa$, $\beta_\kappa/\nu_\kappa$, and $\gamma_\kappa/\nu_\kappa$, respectively. In panel (d), we show that the specific heat $c$ presents a logarithmic divergence with the system size, which yields $\alpha_\kappa=0$.}
\label{ising-regressions}
\end{figure*}

\begin{figure}[H]
\begin{center}
\includegraphics[scale=0.12]{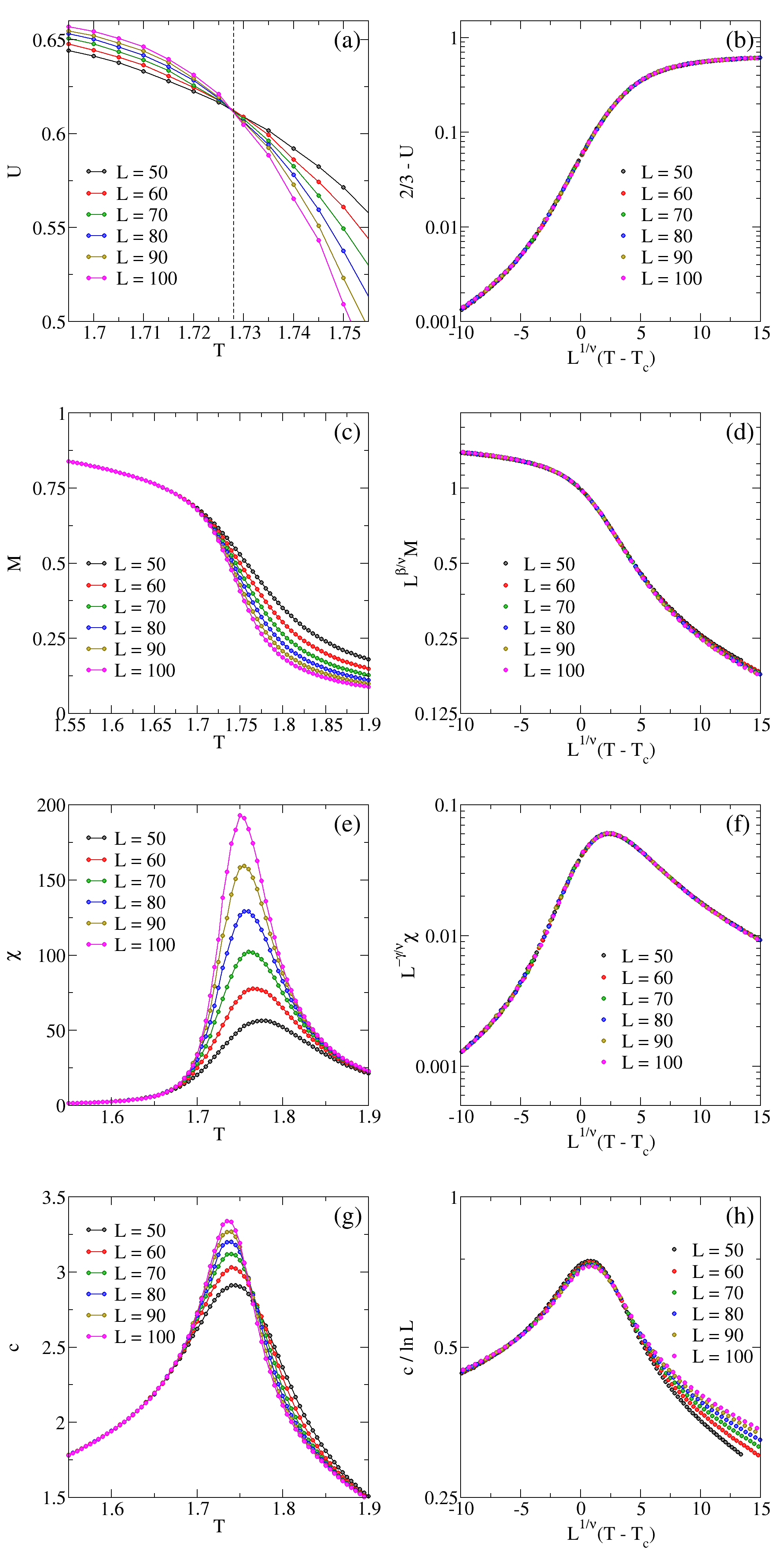}
\end{center}
\caption{We show the Binder cumulant and its data collapse in panels (a) and (b), respectively. The curves for different lattice sizes cross at $T_c \approx 1.728(5)$. In addition, we used the exact 2D Ising exponent $\nu=1$ in the data collapse of the Binder cumulant. We show the magnetization and its data collapse in panels (b) and (c). The data collapse of the magnetization is compatible with the exact 2D Ising exponent $\beta=0.125$. We show the magnetic susceptibility $\chi$ and its data collapse in panels (d) and (e). The magnetic susceptibility diverges with the system size at $T_c$ with the exact 2D ising exponent $\gamma=1.75$. We show the specific heat $c$ and its data collapse in panels (g) and (h). We also note that the specific heat $c$ diverges as $\ln L$, which is compatible with $\alpha=0$. Error bars are smaller than the symbols.}
\label{ising-collapses}
\end{figure}

We show simulation results of the 2D Ising model with the $\kappa$-generalized Metropolis rate in Eq.\ \ref{ising-metropolis-kaniadakis} in Figs.\ \ref{ising-regressions}, and \ref{ising-collapses} for $\kappa=1$. We also simulated $\kappa=0.2$, $\kappa=0.4$, $\kappa=0.6$, and $\kappa=0.8$ (figures not shown), and resume the critical temperatures in Tab.\ \ref{ising-temperatures}. Negative values of $\kappa$ reproduce the curves for respective positive values, the values of the $\kappa$-generalized critical exponents are some even function of $\kappa$. This result is in agreement with that obtained though $\kappa$-SFT, within the margin of error displayed in Table \ref{ising-exponents}. We note $T_c$ eventually vanishes by increasing the $\kappa$ parameter. In addition, in the limit $\kappa \to 0$, we obtain the exact $T_c$ of the Ising model on the square lattice.

\begin{table}[!h]
\caption{Estimates of critical temperatures of 2D Ising model with the $\kappa$-generalized Metropolis rate in Eq.\ (\ref{ising-metropolis-kaniadakis}), for some values of $\kappa$. Estimates of the critical temperature $T_c$ were obtained from the crossings of Binder Cumulant, as seen in panel (a) of Fig.\ \ref{ising-collapses}.}
\begin{tabular}{p{4.0cm}p{2.3cm}}
 \hline
 \hline
 $\kappa$  & $T_c$  \\
 \hline
 1.0   & 1.728(5)   \\
 0.8   & 1.894(5)   \\
 0.6   & 2.041(5)   \\
 0.4   & 2.161(5)   \\
 0.2   & 2.241(5)   \\
 0.001 & 2.269(5)   \\
 \hline
 \hline
\end{tabular}
\label{ising-temperatures}
\end{table}

We can estimate the critical exponent ratio $1/\nu_\kappa$ from the dependence of $d\ln M(T,L)/dT$ on the system size at $T_c$. In addition, we can estimate $\beta_\kappa/\nu_\kappa$, $\gamma_\kappa/\nu_\kappa$, and $\alpha_\kappa/\nu_\kappa$ from the dependence of the magnetization, susceptibility and specific heat on the system size in the critical temperature $T_c$. In Fig.\ \ref{ising-regressions}, we show results of $d\ln M/dT$, $M$, $\chi$ and $c$ as functions of the system size in panels (a), (b), (c), and (d), respectively.

The linear regressions of data in the critical temperature $T_c$ as functions of the system size furnish estimates of the exponent ratios, which we resume in Tab.\ \ref{ising-exponents}. In all results shown in Tab.\ \ref{ising-exponents}, we used $N_\text{term} = 10^6$, and $N_t=10^7$, where we discarded $10^2$ Monte-Carlo steps between two successive elements of the time series. All simulation results are close to the exact 2D Ising model exponents, where we do not observed a strong dependence on $\kappa$.

\begin{table}[!ht]
\caption{Estimates of critical exponent ratios $1/\nu_\kappa$, $\beta_\kappa/\nu_\kappa$, $\alpha_\kappa/\nu_\kappa$ of 2D Ising model with the $\kappa$-generalized Metropolis rate in Eq.\ (\ref{ising-metropolis-kaniadakis}), for some values of $\kappa$, obtained from linear regressions of double log data of the derivative of $\ln M(T, L)$, the magnetization $M$ and the magnetic susceptibility $\chi$ in the critical temperature $T_c$, respectively. All exponent ratios deviate less than $4\%$ from the respective exact values of the 2D Ising model. In all cases, we obtained a logarithmic divergence of $c$, yielding $\alpha_\kappa = 0$.}
\begin{tabular}{p{2.0cm}p{1.6cm}p{1.6cm}p{1.6cm}}
 \hline
 \hline
 $\kappa$    & $1/\nu_\kappa$ & $\beta_\kappa/\nu_\kappa$ & $\gamma_\kappa/\nu_\kappa$ \\
 \hline
 1.0         & 1.023(4)       & 0.1207(7)                 & 1.744(5)                   \\
 0.8         & 1.03(1)        & 0.130(1)                  & 1.78(1)                    \\
 0.6         & 1.03(1)        & 0.125(1)                  & 1.754(4)                   \\
 0.4         & 1.030(5)       & 0.1275(3)                 & 1.764(4)                   \\
 0.2         & 1.02(1)        & 0.128(1)                  & 1.768(8)                   \\
 0.0 (Exact) & 1              & 0.125                     & 1.75                       \\
 \hline
 \hline
\end{tabular}
\label{ising-exponents}
\end{table}

We also show data collapses to confirm the scaling dependence of the thermodynamic properties given in Eq.\ \ref{scaling-functions-ising-2d}, with the exact critical exponents of the 2D Ising model. In all results shown in Fig.\ \ref{ising-collapses}, we used $N_\text{term} = 10^5$, and $N_t=10^7$, where we discarded $10$ Monte-Carlo steps between two successive elements of the time series. We note that the data collapses are not compatible with a strong dependence of the critical exponents on $\kappa$. This deficiency of Kaniadakis statistics could be associated to the fact that it is trace-form but is not composable. For some discussion on the relevance of an entropic functional being simultaneously trace-form and composable, see Ref. \cite{Enciso_2017}.

\par Even if we would consider the just mentioned weak dependence on $\kappa$, we also note that, for some value of $\kappa$, the correspondence between some $\kappa$-generalized critical exponents and the corresponding value of $\kappa$ is not one-two-one. The value of the critical exponents are not uniquely determined for a given value of $\kappa$, \emph{i. e.}, we have two values of $\kappa$ identifying the same value of the critical indices. This is a consequence of the fact that the $\kappa$-generalized distribution is some even function of $\kappa$, namely Eq. (\ref{jhglhjlkisd}). In fact, the set of values of the nonconventional critical exponents for real imperfect crystals, \emph{e. g.} of manganites, can be described for some generalized statistical field theory only if this theory is general and thus furnishes some set of higher and lower values of the critical exponents, when compared to those for perfect crystals \cite{Submitted}. This task is attained only if the distribution is some injective function, namely the case of Ref. \cite{Submitted}, not the present one since an even function is not injective. Then the $\kappa$-generalized statistical field theory is not general and must be discarded as one generalizing statistical mechanics.

\section{Some $\kappa$-generalized models}

\par Although there are not $\kappa$-dependent versions of nongeneralized Ising-like systems, some other $\kappa$-generalized systems exist. We have to present their critical exponents just below.

\subsection{$\kappa$-percolation and $\kappa$-Yang-Lee edge singularity}

\par The $\kappa$-generalized versions, namely both $\kappa$-generalized percolation \cite{Bonfirm_1981} ($\alpha = -1$ and $\beta = -2$) and Yang-Lee edge singularity  \cite{Bonfirm_1981} ($\alpha = -1$ and $\beta = -1$), now for dimensions $d = 6 - \epsilon$, present the following $\kappa$-generalized critical exponents 
\begin{eqnarray}\label{etaphi3}
\eta_{\kappa} = \eta - \frac{4\alpha\beta\kappa^{2}}{3(\alpha - 4\beta)[\alpha - 4\beta(1 - \kappa^{2})]}\epsilon , 
\end{eqnarray}
\begin{eqnarray}\label{nuphi3}
\nu_{\kappa}^{-1} = \nu^{-1} - \frac{20\alpha\beta\kappa^{2}}{3(\alpha - 4\beta)[\alpha - 4\beta(1 - \kappa^{2})]}\epsilon , 
\end{eqnarray}
\begin{eqnarray}\label{omegaphi3}
\omega_{\kappa} = \omega ,
\end{eqnarray}
where $\eta_{\kappa}$ and $\nu_{\kappa}$ are the corresponding nongeneralized critical exponents up to all-loop order. As \emph{i. e.} $1 < \kappa < 1$. For Yang-Lee edge singularity, $\eta_{\kappa}$ and $\nu_{\kappa}$ are not independent \cite{Bonfirm_1981,PhysRevD.95.085001} and they are related by $\nu_{\kappa}^{-1} = (d - 2 + \eta_{\kappa})/2$ \cite{Bonfirm_1981,PhysRevD.95.085001}. From $\eta_{\kappa}$ and $\alpha = -1$ and $\beta = -1$ we compute $\nu_{\kappa}$. We can evaluate the remaining $\kappa$-generalized critical indices from the scaling relations among them \cite{PhysRevD.103.116024}.

\subsection{$\kappa$-long-range $\lambda\phi^{3}$ theory}
\par For the long-range $\lambda\phi^{3}$ theory \cite{PhysRevB.31.379} in $d = 3\sigma - \varepsilon$ the corresponding $\kappa$-generalized critical exponents can be written as  
\begin{eqnarray}
\eta_{\sigma , \hspace{.5mm}\kappa} = \eta_{\sigma}, \hspace{1cm} \nu_{\sigma, \hspace{.5mm}\kappa}^{-1} = \nu_{\sigma}^{-1} - \frac{\kappa^{2}}{1 - \kappa^{2}}\frac{\alpha}{2\beta}\epsilon ,
\end{eqnarray}
where $\alpha$ and $\beta$ assume the values $-1$, $-1$ and $-1$, $-2$ for the Yang-Lee edge singularity problem and percolation cases \cite{Bonfirm_1981}, respectively. The nongeneralized value of $\eta_{\sigma} = 2 - \sigma$ is exact \cite{PhysRevB.31.379} and $\eta_{\sigma ,\hspace{.5mm}\kappa}$ is exact within the approximation of this work. The nongeneralized exponents values were obtained up to two-loop level in the earlier work \cite{PhysRevB.31.379}.

\subsection{$\kappa$-Gross-Neveu-Yukawa model}
\par The $\kappa$-generalized Gross-Neveu-Yukawa model \cite{ZINNJUSTIN1991105} expresses interacting scalar field $\phi$ and $N$ massless Dirac fermions $\psi$ and $\bar{\psi}$ in $d = 4 - \epsilon$ dimensions. The corresponding $\kappa$-generalized critical indices are given by \cite{ZinnJustin} 
\begin{eqnarray}
\eta_{\psi ,\hspace{.5mm}\kappa} = \eta_{\psi} + \frac{\kappa^{2}}{(2N + 3)(2N + 3 - 2\kappa^{2})}\epsilon , 
\end{eqnarray}
\begin{eqnarray}
\eta_{\phi ,\hspace{.5mm}\kappa} = \eta_{\phi} + \frac{4N\kappa^{2}}{(2N + 3)(2N + 3 - 2\kappa^{2})}\epsilon ,
\end{eqnarray}
\begin{eqnarray}
\nu_{\kappa}^{-1} = \nu^{-1} - \frac{A_{N,\hspace{.5mm}\kappa}}{(2N + 3)(2N + 3 - 2\kappa^{2})}\epsilon ,
\end{eqnarray}
where 
\begin{eqnarray}
A_{N,\hspace{.5mm}\kappa} = (2N + 3)(R_{N,\hspace{.5mm}\kappa}/6 + 2N) - (2N + 3 - 2\kappa^{2})(R_{N}/6 + 2N),
\end{eqnarray}
\begin{eqnarray}
R_{N,\hspace{.5mm}\kappa} = -(2N - 3 + 2\kappa^{2}) + \sqrt{(2N - 3 + 2\kappa^{2})^{2} + 144N(1 - 4\kappa^{2})},
\end{eqnarray}
\begin{eqnarray}
R_{N} = \lim_{\kappa\rightarrow 0} R_{N,\hspace{.5mm}\kappa}.
\end{eqnarray}
The nongeneralized critical exponents $\eta_{\psi}$, $\eta_{\phi}$ and $\nu$ were evaluated up to four-loop level in Ref. \cite{PhysRevD.96.096010}.

\subsection{$\kappa$-short- and $\kappa$-long-range directed percolation}
\par For $\kappa$-generalized short- and long-range directed percolation \cite{JANSSEN2005147,Tauber_2005} in $d = 4 - \epsilon$ and $d = 2\sigma - \varepsilon$, respectively, we obtain 
\begin{eqnarray}
\eta_{\kappa} = \eta - \frac{\kappa^{2}}{1 - \kappa^{2}}\frac{\epsilon}{6}, \hspace{1cm} \eta_{\sigma ,\hspace{.5mm}\kappa} = \eta_{\sigma} - \frac{\kappa^{2}}{1 - \kappa^{2}}\frac{\varepsilon}{7},
\end{eqnarray}
\begin{eqnarray}
\nu_{\kappa} = \nu + \frac{\kappa^{2}}{1 - \kappa^{2}}\frac{\epsilon}{16}, \hspace{1cm} \nu_{\sigma ,\hspace{.5mm}\kappa} = \nu_{\sigma} + \frac{\kappa^{2}}{1 - \kappa^{2}}\frac{2\varepsilon}{7\sigma^{2}},
\end{eqnarray}
\begin{eqnarray}
z_{\kappa} = z - \frac{\kappa^{2}}{1 - \kappa^{2}}\frac{\epsilon}{12}, \hspace{1cm} z_{\sigma ,\hspace{.5mm}\kappa} = z_{\sigma} - \frac{\kappa^{2}}{1 - \kappa^{2}}\frac{\varepsilon}{7}. 
\end{eqnarray}
The nongeneralized indices were computed up to two-loop level in \cite{JANSSEN2005147}.

%

\subsection{$\kappa$-short- and $\kappa$-short-range dynamic isotropic percolation}

\par In the case of $\kappa$-generalized short- and long-range dynamic isotropic percolation \cite{JANSSEN2005147,Tauber_2005} at $d = 6 - \epsilon$ and $d = 3\sigma - \varepsilon$, respectively, we have  
\begin{eqnarray}
\eta_{\kappa} = \eta - \frac{\kappa^{2}}{1 - \kappa^{2}}\frac{\epsilon}{21}, \hspace{1cm} \eta_{\sigma ,\hspace{.5mm}\kappa} = \eta_{\sigma} - \frac{\kappa^{2}}{1 - \kappa^{2}}\frac{3\varepsilon}{8}, 
\end{eqnarray}
\begin{eqnarray}
\nu_{\kappa} = \nu + \frac{\kappa^{2}}{1 - \kappa^{2}}\frac{5\epsilon}{84}, \hspace{1cm} \nu_{\sigma ,\hspace{.5mm}\kappa} = \nu_{\sigma} + \frac{\kappa^{2}}{1 - \kappa^{2}}\frac{\varepsilon}{4\sigma^{2}},
\end{eqnarray}
\begin{eqnarray}
z_{\kappa} = z - \frac{\kappa^{2}}{1 - \kappa^{2}}\frac{\epsilon}{6}, \hspace{1cm} z_{\sigma ,\hspace{.5mm}\kappa} = z_{\sigma} - \frac{\kappa^{2}}{1 - \kappa^{2}}\frac{3\varepsilon}{16}.
\end{eqnarray}
The nongeneralized critical indices were computed up to two-loop level in Ref. \cite{JANSSEN2005147}.

%

\section{Physical interpretation of the results}

\par The physical interpretation of the theory can be seen, \emph{e. g.}, from the results for the critical indices for both $\kappa$-percolation and $\kappa$-Yang-Lee edge singularity shown in Tables \ref{tableexponentsd2p}-\ref{tableexponentsd2LY} just below

\begin{table}[h]
\caption{Exact $\kappa$-generalized critical exponents, for some values of $\kappa$, to $2$d $\kappa$-generalized percolation, obtained from $\kappa$-SFT.}
\begin{tabular}{ p{4.0cm}p{2.3cm}p{1.1cm}  }
 \hline
 \hline
 $\kappa$  & $\beta_{\kappa}$ & $\gamma_{\kappa}$   \\
 \hline
 0.4  &  0.155 & 3.417    \\
 0.3  &  0.145 & 2.832    \\
 0.2  &  0.141 & 2.557    \\
 0.1  &  0.139 & 2.427    \\
 0.0 Exact\cite{PhysRevD.103.116024}  &  0.139  &  2.389          \\
 \hline
 \hline
\end{tabular}
\label{tableexponentsd2p}
\end{table}

\begin{table}[h]
\caption{Exact $\kappa$-generalized critical exponents, for some values of $\kappa$, to $2$d $\kappa$-generalized Yang-Lee edge singularity, obtained from $\kappa$-SFT.}
\begin{tabular}{ p{4.0cm}p{2.3cm}p{1.1cm}  }
 \hline
 \hline
 $\kappa$  & $\beta_{\kappa}$ & $\gamma_{\kappa}$   \\
 \hline
 0.8  &   1.001  &  -3.183    \\
 0.6  &   1.000  &  -5.306    \\
 0.4  &   1.000  &  -6.344    \\
 0.2  &   1.000  &  -6.848    \\
 0.0 Exact\cite{PhysRevD.103.116024}  &  1.000  &  -7.000          \\
 \hline
 \hline
\end{tabular}
\label{tableexponentsd2LY}
\end{table}

\par We observe that the $\kappa$-generalized critical indices numerical values turn out to be higher than the nongeneralized one when $\kappa$ ranges away the nongeneralized value $\kappa = 0$ (both for $\kappa > 0$ and $\kappa < 0$ due to the fact that the $\kappa$-generalized exponential function is an even function of $\kappa$). Then now we present the physical interpretation of such results: from their definitions, the critical exponents furnish a measure of how much a given physical quantity diverges near the system critical point. In the case, for example, of the inverse susceptibility of a given material, we can obtain information about how much the system is susceptible to changes in the magnetic field. So the susceptibility diverges stronger (weaker) than in the nongeneralized case when the corresponding critical index, namey $\gamma$, displays higher (lower) numerical values. Then higher (lower) numerical values of the critical exponents means more (less) susceptible systems to magnetic field changes and thus systems interacting weakly (strongly) than the nongeneralized situation. Now the $\kappa$-parameter can be physically interpreted as one encoding some effective weaker interaction than in the nongeneralized case. Alternatively, we can predict the behavior of the system (with the energy $E < 0$ in units of $k_{B}T$) from 
\begin{eqnarray}
e_{\kappa}^{-E} \approx e^{-E}\left(1 + \frac{1}{6}\kappa^{2}E^{3}\right) \approx e^{-\left(E - \frac{1}{6}\kappa^{2}E^{3}\right)}.
\end{eqnarray}
In the approximation aforementioned, we have the effective energy ($E < 0$) $E - \frac{1}{6}\kappa^{2}E^{3}$. It increases or gets weaker (never decreases or gets stronger since the $\kappa$-generalized exponential function is an even function of $\kappa$) for all values of $\kappa$ through its range. As the effective energy or interaction always turn out to be weaker, the system must have $\kappa$-generalized critical indices higher than that in the nongeneralized situation as can be seen in Tables \ref{tableexponentsd2p}-\ref{tableexponentsd2LY}. Furthermore, as the effective $\kappa$-generalized energy never decreases or presents stronger values, we can not never obtain $\kappa$-generalized critical exponents with smaller numerical values when compared with the nongeneralized one. As there are many real materials for which the corresponding critical exponents values are smaller that the nongeneralized one, the critical behavior of these materials can not be explained by applying the $\kappa$-generalized distribution, thus characterizing such a distribution as some incomplete one, as it was done by using the nonextensive one \cite{Submitted}.

\section{Conclusions}

\par We have introduced some general field-theoretic approach for studying the critical properties of systems undergoing continuous phase transitions in the $\kappa$-generalized statistics framework, namely $\kappa$-generalized statistical field theory. We have showed that some supposedly $\kappa$-generalized systems, \emph{e. g.} $\kappa$- Ising, Heinsenberg, $\phi^{6}$, long-range, Gross-Neveu, uniaxial strong dipolar forces, spherical, Lifshitz and multicritical models do not present the same behavior as their nongeneralized counterparts. It it is not suitable for describing the nonconventional critical properties of real imperfect crystals, \emph{e. g.} of manganites, as some alternative generalized theory is, namely nonextensive statistical field theory, as shown recently in literature. This implies that the $\kappa$-generalized statistical field theory is not general and must be discarded as one generalizing statistical mechanics. Although $\kappa$-generalized versions of the systems aforementioned do not exist, we have displayed a few ones that depend on $\kappa$, for which we have presented the corresponding physical interpretation through the general physical interpretation of the $\kappa$-parameter.

\section*{Declaration of competing interest}

\par The authors declare that they have no known competing financial interests or personal relationships that could have appeared to influence the work reported in this paper.

\section*{Acknowledgments}

\par PRSC would like to thank the Brazilian funding agencies CAPES and CNPq (grants: Universal-431727/2018 and Produtividade 307982/2019-0) for financial support.

\bibliography{apstemplate}

\end{document}